\documentclass[12pt,preprint]{aastex}

\newcommand{\xmm}{{\it XMM-Newton}}

\newcommand{\chandra}{{\it Chandra}}

\newcommand{\hetgs}{{\small HETGS}}

\newcommand{\astroe}{{\it ASTRO-E2}}

\newcommand{\agn}{{\small AGN}}
\newcommand{\uta}{{\small UTA}}

\newcommand{\hullac}{{\small HULLAC}}

\newcommand{\Uox}{{$U_{Ox}$}}  
\newcommand{\cm}{{\rm\thinspace cm}} 
\newcommand{\cmii}{\hbox{$\cm^{-2}\,$}}
\newcommand{\cmiii}{\hbox{$\cm^{-3}\,$}}                                                                             
\newcommand{\mcg}{MCG--6-30-15}
\newcommand{\iras}{IRAS 13349+2438}

\shorttitle{Inner-shell absorption lines}
\shortauthors{Behar \& Netzer}

\begin{document}

\title{Inner-shell 1s - 2p Soft X-ray Absorption Lines}
\author{Ehud Behar \altaffilmark{1}, Hagai Netzer \altaffilmark{2}}

\altaffiltext{1}{Columbia Astrophysics Laboratory and Department of Physics,
                 Columbia University, 550 West 120th Street, New York, NY
                 10027; behar@astro.columbia.edu}
\altaffiltext{2}{School of Physics and Astronomy, Raymond and Beverly Sackler
Faculty of Exact Sciences, Tel-Aviv University, Tel-Aviv 69978, Israel}

\received{11/16/2001}
\revised{}
\accepted{}

\shorttitle{1s - 2p X-ray absorption}
\shortauthors{Behar \& Netzer}

\begin{abstract}
The \hullac\ atomic code is used to compute wavelengths and oscillator
strengths for the 1s - $n$p inner-shell absorption
lines in Li-like to F-like ions of neon, magnesium, aluminum, silicon, sulfur,
argon, calcium, and iron. Many of these lines are expected to be observed in
\chandra\ and \xmm\ high-resolution X-ray spectra of active galaxies. The new atomic
data are incorporated in the ION code for spectral modeling of photoionized
plasmas. The calculated spectra are subsequently compared with the spectrum of NGC~3783 and show very good agreement. The usefulness of
these lines as diagnostics for the ionization state, column densities, and
velocities in line-of-sight photoionized gas is called attention to. 

\end{abstract}


\section{INTRODUCTION}
\label{sec:intro}
The launch of the \chandra\ and \xmm\ observatories along with the grating
spectrometers on board has generated great interest in high resolution
X-ray spectroscopy of a wide variety of astrophysical X-ray sources, in
particular active galactic nuclei (\agn). The high resolution grating
observations of a handful of \agn, e.g. NGC~5548  \citep {kaastra00}, NGC~3783
\citep {kaspi00, kaspi01}, \mcg\ \citep {br01, lee01}, and \iras\ \citep {sako01}, clearly show the presence of dozens of
strong X-ray absorption lines originating from highly charged gas along the
line of sight. While many of the absorption lines are due to transitions in H-like and He-like ions, some
are clearly due to inner-shell transitions in lower ionization species. 
Perhaps the most prominent feature of this kind is the unresolved transition array (\uta)
of inner-shell $n$~= 2~to~3 (mainly 2p~- 3d) lines pertaining to various
M-shell Fe
ions identified in \iras\ \citep {sako01}. The atomic data needed to model this \uta\ feature have recently been
published in \citet {behar01}. Another example is the inner-shell 1s - 2p feature
of Li-like Si$^{11+}$  in NGC 3783 reported by \citet {kaspi01}. The analogous
1s - 2p lines for Li-like O$^{5+}$ have been calculated by \citet {pradhan00}
using the R-matrix method and identified in \mcg\ by \citet {lee01}. \citet
{sako02} have found evidence for additional 1s - $n$p lines arising from lower charge states
of oxygen (O$^{4+}$ and O$^{3+}$) in \mcg. Recently, \citet {nahar01}
published results for Li-like C$^{3+}$, O$^{5+}$, and Fe$^{23+}$ obtained
using the same method as \citet {pradhan00}.

Models like
those presented in \citet {kaspi01} predict that inner-shell transitions from
the ground level of many more L-shell ions (Li-like to F-like) are indeed
expected for X-ray illuminated photoionized plasmas. However, the available
atomic data are few. The present paper aims at providing comprehensive results for the resonant 1s - $n$p transitions in Li-like through
F-like ions of neon, magnesium, aluminum, silicon, sulfur, argon, calcium, and
iron. For that purpose, we employ the \hullac\ code, which uses the
parametric potential approximation. This method is by far more efficient than the previously used methods. For the present case of inner-shell transitions
in highly charged ions, it is also expected to be very accurate. In the
following, \S2 describes the method of calculation and the new
atomic data and \S3 shows theoretical models for \agn\ spectra and a
comparison with the spectrum of NGC~3783 obtained with the  \hetgs\
spectrometer on board \chandra.

\section{ATOMIC CALCULATIONS}
\label{sec:atomic_calc}
In order to calculate the wavelengths and oscillator strengths of interest, we use the 
multi-configuration, relativistic \hullac\
(Hebrew University Lawrence Livermore Atomic Code) computer package \citep
{bs01}. The level energies in \hullac\ are calculated using the
relativistic version of the Parametric Potential (PP) method by \citet
{klapisch77}. This relativistic approach is perfectly adequate for
the high-energy inner-shell excitations calculated in the present work. The high efficiency of
\hullac\ allows us to calculate the numerous transitions involved relatively
quickly. Unlike some other atomic methods, the PP approach is hardly
hampered by the considerable complexity of the electronic structure in the
ions considered. Calculations are carried out
for 1s - $n$p ($n \le 7$) transitions in the L-shell ions (Li-like to F-like)
of Ne, Mg, Al,
Si, S, Ar, Ca, and Fe. The atomic structure calculations include the ground
configurations 1s$^2$2$l^x$ ($x$~= 1 to 7). For each ion, all of the 1s - $n$p
photoexcitation transitions from the ground level are calculated,
among which the 1s - 2p are the strongest by far. Electric and magnetic dipole
and quadrupole transitions are computed. The contributions, however, of
excitations other than electric-dipole are negligible.

The calculated 1s - $n$p wavelengths ($\lambda _{ij}$) and oscillator
strengths ($f_{ij}$) for the strongest absorption lines in the He~I to F~I
isoelectronic sequences are listed in Table 1. Only lines with $f_{ij} > 0.1$,
or the strongest line for a given ion, are included in the table. These lines
represent the major part ($>$70\%, and in most cases $>$90\%) of the
absorption effect for the respective transitions. Thus, they should
be sufficient for the analysis of most astrophysical spectra. We note,
however, that the ION code, for which results are shown below does include
more lines than those presented in Table~1. The strong lines are predominantly
1s - 2p lines, but also 1s - 3p lines in the He~I and Be~I sequences (see table). 
More specifically, the strongest lines usually arise for 1s -
2p$_{3/2}$ transitions.  The Einstein coefficient ($A_{ji}$) as well as the 
total rate for depletion of the inner-shell excited level
$[\Sigma (A+A^a)]$, which includes both radiative decay ($A$) and
autoionization ($A^a$) rates, are also given in Table~1. The total depletion rate is needed to
compute the fluorescence yield  $A_{ji}~/~\Sigma(A+~A^a)$ and natural width
associated with each absorption line.

As can be seen in Table~1, and as expected, the oscillator strength depends
only very weakly on the element and is much more contingent upon the
isoelectronic sequence. Table~1 shows that for each element, the 
oscillator strength generally increases with the ionic charge, as more 
vacancies become available in the 2p subshell. This trend appears to be 
violated by the lines of the Be-like ions, which feature a 1s - 2p line
that is stronger than the lines of the Li-like ions. 
However, the total oscillator strength, summed over all of the
1s - 2p excitations for Li-like and Be-like species is in fact comparable,
as expected for these systems, both of which have the maximum possible six
vacancies in the 2p subshell. The apparent dominance of the Be I sequence is
only due to the fact that for the Be-like ions, the entire oscillator strength
is concentrated in one 1s - 2p$_{3/2}$ line. The same stands for the 1s -
3p$_{3/2}$ transition (as well as for higher $n$), which explains why those transitions in
Be-like ions are strong enough ($f_{ij} > 0.1$) to make it into the
table. Conversely, for the Li-like ions the total 1s - 2p$_{3/2}$ line strength is distributed
among several lines.

The accuracy of all of the data presented here is expected to be very high,
but owing to the relativistic method in \hullac, the data become increasingly
accurate for higher charge states. In order to give a rough idea of the
uncertainties that could be expected from these atomic values, we include
in Table~1, in addition to the new inner-shell excited lines, the leading lines
for the He-like ions, for which all of the data are available in the literature. The
\hullac\ wavelengths and oscillator strengths for
the 1s - $n$p ($n$~= 2, 3) excitations in He-like ions are compared in the
table with the published values, which were calculated with perhaps more rigorous
methods. For this comparison we use primarily the compilation by \citet
{verner96} as well as values from the NIST database,
in cases where these are available (see Table~1). It is seen in Table~1 that the present
wavelengths agree with the published values to within a few m\AA\ and that the
oscillator strengths agree to within $\sim$10\% (except for Fe in \citet {verner96}).
Similar comparisons for the {\it inner-shell} excitations in Table~1
would, of course, be even
more desirable. Unfortunately, calculations for these transitions have
been previously published only for Li-like Fe$^{23+}$. \citet {nahar01} obtain
for the two strongest lines of Fe$^{23+}$ 1.860 \AA\ and 1.864 \AA, to be
compared with the present values of 1.861 \AA\ and 1.864 \AA. The
corresponding oscillator strengths in \citet {nahar01} are 0.491 and 0.146, to
be compared with the present values of 0.469 and 0.147. In order to check
whether this excellent agreement is sustained for less ionized species, we have
specifically calculated the 1s - 2p lines for the Li-like ion O$^{5+}$, which is
beyond the scope of this work and therefore not included in Table~1. Our
calculations show that the two strongest 1s - 2p lines of O$^{5+}$ overlap at
22.01 \AA\ and have oscillator strengths of 0.351 and 0.174. These values
compare favorably with \citet {nahar01}, who find the lines (overlapping)
at 22.05 \AA, and the oscillator strengths to be 0.384 and 0.192. The agreement between the
autoionization rates obtained with the two methods is of the same order as that
for the oscillator strengths. In other
words, even for the O$^{5+}$ case, which could be considered an extreme upper limit to
the uncertainty expected from the present values for the higher-Z elements, 
the discrepancies with the results obtained with the more rigorous R-matrix code
are not more than 0.2\% for the wavelengths and 10\% for the oscillator
strengths and autoionization rates. The accuracy of most of the present data
is, however, expected to
be much higher as indicated by the comparisons for the He-like ions and for Fe$^{23+}$.
Having gone through this extensive comparison, it is important to
stress that for an ultimate assessment of the accuracy of the atomic data,
a comparison between calculations may not be sufficient. Actual measurements
need to be carried out.

With the exception of maybe Fe, the lines from each ion
are adequately separated in wavelength space and are readily resolvable with
the gratings on board \chandra\ and \xmm. The uncertainties associated with
the wavelengths calculated with \hullac\ are considerably smaller than the
typical differences in wavelength. This enables an unambiguous identification
of lines and, consequently, a precise measurement of the ionization state of the absorbing plasma. We note that although the inner-shell Fe lines are
within the wavelength band of the \hetgs\ spectrometer on board \chandra\,
they are difficult to resolve with \hetgs. These lines will be much better
detected and resolved with the microcalorimeter spectrometer \citep
{richard99}, which is planned to fly on board \astroe.

For most of the inner-shell excited levels considered in this work, the total
autoionization rate is much higher than the radiative decay rate, especially
for the low-$Z$ elements. The fact
that the inner-shell photoexcitation processes are by and large followed by
autoionization rather than X-ray re-emission is manifested in the generally
low fluorescence yield, which gradually increases with the ion
charge. For example, for Si and Fe, respectively, the $f$-weighted averages of
the fluorescence yield rise gradually from $\sim$ 0.04 and 0.25 for the F-like 
ion to $\sim$ 0.27 and 0.93 for the Li-like ion.
The low fluorescence yield can be associated with several important
effects. First, autoionization following photoexcitation can alter the
ionization balance towards higher charge states. This effect is
included in ION and can be important (see \S3). Another
direct result of the low fluorescence yield is that dielectronic recombination 
via these channels is inefficient and does not compensate for the
photoexcitation-autoionization effect on the ionization balance.  It further implies
that these transitions are much harder to detect in emission than they are in
absorption. 
Another consequence of the high autoionization rates is the broadening of the 
natural widths of the inner-shell absorption lines. At high column densities, 
the natural width of the line becomes important for determining the total flux absorbed.

\section{SPECTRAL MODELING}
\label{sec:modeling}
We have carried out detailed spectral modeling in order to illustrate the expected
 intensity of the inner-shell transitions and their use as diagnostics for
photoionized plasma. The calculations are performed with ION2001, the 2001 version of the photoionization
code ION lately described in \citet {netzer01},  \citet {kaspi01}, and  \citet
 {netzer96}. The ION code
computes a self-consistent model of the ionization and thermal structure
of a steady-state ionized gas exposed to an external radiation source. Atomic
data include the most recent cross sections for the more abundant elements
and all of their strong lines.
Also included are $f$-values calculated with \hullac\ for a large number of iron L-shell lines
(see  \citet {kaspi01}) and dozens of M-shell iron transitions \citep {behar01}.  The new inner-shell transitions calculated in the present work are fully incorporated into the model,
including their effect on the individual level population and on the level of
 ionization of the various elements due to photoexcitation-autoionization. 

We have calculated a series of AGN-type models with a column density of 10$^{22}$ \cmii, 
gas density of 10$^8$ \cmiii, and a range of ionization parameters similar to the one observed
in many Seyfert galaxies. 
The local ionizing field is defined by the spectral energy distribution (SED)
of the ionizing sources, taken here to be a single power-law of photon slope
$\Gamma=1.8$ between 0.1 to 50 keV, and by the oxygen ionization parameter,
\Uox, defined over the 0.538 -- 10 keV range \citep {netzer01}.  Guided by recent high-resolution \chandra\ observations, we have assumed a turbulent velocity
of 300 km/s for the absorbing gas. Solar elemental composition is assumed in all cases.

Fig. 1 shows the 4.7 -- 10 \AA\ wavelength range for three pure-absorption
theoretical spectra calculated for \Uox: 10$^{-1.5}$, 10$^{-2}$, and 10$^{-2.5}$. The largest optical-depth transitions for each of
the H-like to F-like ions are marked above the plots. As evident from this example, a relatively
small change in incidence flux makes a large change in the level of ionization
and, consequently, in the spectral line structure. 
The 1s - 2p inner-shell lines of at least 2 - 4 L-shell ions arise for a single
ionization parameter, and therefore are likely to be detected in
many \agn\ absorption spectra. 
This, in combination with the observed H-like and He-like lines, provides a
powerful diagnostic tool for determining very precisely the level of
ionization and the elemental abundances in the absorbing gas. 
The strongest lines in all cases have typical equivalent widths of several
m\AA\ and are, thus, easily detected in typical \agn\ spectra. The ample
separation between lines from different charge states and the very few blends
allow the identification and reliable curve-of-growth analysis for all of
these lines. Thus, in addition to the ionization state, the present atomic
data enable accurate column density and velocity broadening measurements with
these intrinsically narrow features. In fact, the major remaining limitation
on velocity measurements is the resolution of contemporary airborne, X-ray
spectrometers (typically, a few 100 km/s).

The effect of the autoionization transitions following photoexcitation on the
level of ionization has been tested for the \Uox=10$^{-2}$ case by running ION
models with and without the inner-shell lines. We find a noticeable change
in the level of ionization near the illuminated face of the cloud, where most
of the absorption is expected. For example, including the inner-shell
transitions changes the fractional abundance of Si$^{10+}$ from 0.21 to 0.17,
that of Si$^{11+}$ from 0.27 to 0.25, and that of Si$^{12+}$ from 0.35 to
0.43, i.e.,  silicon becomes appreciably more ionized when
photoexcitation-autoionizaion is included in the model. However, the effect on
 the mean level of ionization deeper in the cloud would be smaller, since the 
strong 1s - 2p lines saturate, which results in photoexcitation being less 
efficient.

The present calculations can be compared with recently obtained spectra of
active galaxies. In particular, the \citet {kaspi01} observation of NGC~3783
is a good example of a photoionized line-of-sight gas with a line-rich absorption spectrum and with properties that are similar to the ones 
investigated here. In \citet {kaspi01}, we tentatively identified features
near 6.8 \AA\ to be due to inner-shell absorption in Si. However, the
atomic data available at that time did not allow for a conclusive
identification. Using the newly calculated atomic data, together with our
spectral code, we now revisit the \citet {kaspi01} observation. In
particular, the present model is compared with the data in the wavelength
region suspected for showing inner-shell Si lines.
The results are shown in Fig. 2 for two representing ionization
parameters. Each absorption trough in the spectrum is attributed for the most
part to a single charge state. Absorption lines of He-like Si$^{12+}$ to
N-like Si$^{7+}$ are clearly identified in the spectrum of NGC 3783.
It can be seen that the \Uox=10$^{-2}$ model reproduces the Si$^{8+}$ -
Si$^{11+}$ lines fairly well; both the wavelengths and the equivalent
widths. Fig. 2 also confirms the \citet {kaspi01} finding that
the data require a range of ionization parameters. In particular,
a component with \Uox $> 10^{-2}$ is required to explain the Si$^{12+}$ line
at 6.65 \AA, and one with \Uox $< 10^{-2}$ might be needed to account for the
Si$^{7+}$ feature at 7.00 \AA. We are currently working on a more detailed fit, including lines
from all of the other cosmically abundant elements to fit the new 900 ks
\chandra\ \hetgs\ data set obtained recently \citep {kaspi02, george02}.

\section{CONCLUSIONS}
\label{sec:disc}
A complete set of wavelengths and oscillator strengths for inner-shell
absorption lines are calculated for the entire L-shell of Ne, Mg, Al, Si, S, Ar, Ca, and Fe. This is a
continuation of our ongoing study of inner-shell absorption features in \agn\
spectra. The usefulness of these atomic data for plasma diagnostics is
demonstrated by comparison of spectral models with the \hetgs\ line-rich
spectrum of NGC 3783. Good agreement is found between the model and the data,
although for a full account of the X-ray absorber in NGC 3783, the \hetgs\
data clearly require a multi ionization-parameter model, which will be published separately.

\acknowledgments

We are grateful to Steven Kahn, Masao Sako and Shai Kaspi for useful
discussions. We thank the referee for suggesting that we include the He-like
ions in Table~1 to demonstrate the accuracy of the present method. 
HN acknowledges helpful discussions with Anil Pradhan and
support by the Israel Science Foundation and the Jack Adler Chair of
Extragalactic Astronomy at Tel Aviv University. HN thanks the Astrophysics
Laboratory group at Columbia University for their hospitality and support
during the period when this research was conducted.

\begin{deluxetable}{lccccccc}
%\rotate
%\tabletypesize{\scriptsize}
%\tablecolumns{6}
\tablewidth{0pt}
\tablecaption{Strongest 1s - $n$p absorption lines from the ground level of
He-like to F-like ions. \label{tab1}}
\tablehead{
  \colhead{Element} &
  \colhead{Isoelectronic} &
  \multicolumn{2}{c}{$\lambda_{ij}$ (\AA) } &
  \multicolumn{2}{c}{$f_{ij}$} &
  \colhead{$A_{ji}$  } &
  \colhead{$\Sigma (A+A^a)$ } \\
& Sequence & present & previous\tablenotemark{a}& present & previous\tablenotemark{a}& (10$^{13}$~s$^{-1}$)& (10$^{13}$~s$^{-1}$)}
\startdata

Ne & He I & 13.448 & 13.447 & 0.657 & 0.724 & 0.81 & 0.81 \\
   &      & 11.553 \tablenotemark{b} & 11.547 & 0.159 & 0.149 & 0.27 & 0.28 \\
   & Li I & 13.646 & & 0.381 & & 0.68 & 1.30 \\
   &      & 13.648 & & 0.187 & & 0.67 & 1.43 \\
   & Be I & 13.814 & & 0.595 & & 0.69 & 8.25 \\
   &      & 12.175 \tablenotemark{b}& & 0.126 & & 0.19 & 0.21 \\
   & B I  & 14.020 & & 0.190 & & 0.64 & 3.55 \\
   &      & 14.047 & & 0.166 & & 0.28 & 18.4 \\
   & C I  & 14.239 & & 0.147 & & 0.16 & 24.9 \\
   &      & 14.202 & & 0.124 & & 0.14 & 5.76 \\
   & N I  & 14.371 & & 0.119 & & 0.26 & 27.3 \\
   & O I  & 14.526 & & 0.106 & & 0.34 & 25.5 \\
   & F I  & 14.631 & & 0.062 & & 0.39 & 0.58 \\
Mg & He I &  9.170  & 9.169 & 0.691 & 0.742, 0.745 \tablenotemark{c}& 1.81 & 1.81 \\
   &      &  7.854 \tablenotemark{b} & 7.851 & 0.159 & 0.151, 0.152 \tablenotemark{c} & 0.57 & 0.61 \\
   & Li I &  9.281  & & 0.402 & & 1.56 & 2.12 \\
   &      &  9.283  & & 0.194 & & 1.50 & 2.36 \\
   & Be I &  9.378  & & 0.631 & & 1.60 & 9.67 \\
   &      &  8.201 \tablenotemark{b} & & 0.132 & & 0.44 & 0.46 \\
   & B I  &  9.498  & & 0.207 & & 1.53 & 5.08 \\
   &      &  9.514  & & 0.188 & & 0.69 & 20.8 \\
   & C I  &  9.631  & & 0.166 & & 0.40 & 29.1 \\
   &      &  9.607  & & 0.143 & & 0.34 & 7.95 \\
   & N I  &  9.718  & & 0.132 & & 0.62 & 32.5 \\
   & O I  &  9.816  & & 0.125 & & 0.86 & 35.7 \\
   & F I  &  9.895  & & 0.076 & & 1.04 & 41.1 \\
Al & He I &  7.757 & 7.757 & 0.693 & 0.750 & 2.56 & 2.56 \\
   &      &  6.637 \tablenotemark{b} & 6.635 & 0.159 & 0.152, 0.153 \tablenotemark{c}& 0.80 & 0.85 \\
   & Li I &  7.844  & & 0.410 & & 2.22 & 2.76 \\
   &      &  7.486  & & 0.196 & & 2.12 & 3.05 \\
   & Be I &  7.921  & & 0.645 & & 2.28 & 10.6 \\
   &      &  6.907 \tablenotemark{b} & & 0.134 & & 6.25 & 0.64 \\
   & B I  &  8.016  & & 0.214 & & 2.22 & 6.19 \\
   &      &  8.029  & & 0.198 & & 1.03 & 21.9 \\
   & C I  &  8.121  & & 0.176 & & 0.59 & 30.8 \\
   &      &  8.101  & & 0.152 & & 0.52 & 9.58 \\
   & N I  &  8.193  & & 0.137 & & 0.91 & 34.8 \\
   & O I  &  8.272  & & 0.131 & & 1.28 & 40.2 \\
   & F I  &  8.337  & & 0.082 & & 1.57 & 47.7 \\
Si & He I &  6.648 & 6.648 & 0.701 & 0.757 & 3.53 & 3.53 \\
   &      & 5.682 \tablenotemark{b} & 5.681 & 0.159 & 0.152 & 1.09 & 1.16 \\
   & Li I &  6.717  & & 0.418 & & 3.09 & 19.9 \\
   &      &  6.719  & & 0.197 & & 2.90 & 5.51 \\
   & Be I &  6.778  & & 0.656 & & 3.17 & 29.6 \\
   &      &  5.896 \tablenotemark{b} & & 0.136 & & 0.87 & 4.41 \\
   & B I  &  6.854  & & 0.221 & & 3.14 & 26.8 \\
   &      &  6.864  & & 0.209 & & 1.48 & 27.6 \\
   & C I  &  6.939  & & 0.185 & & 0.86 & 33.0 \\
   &      &  6.923  & & 0.161 & & 0.75 & 37.2 \\
   & N I  &  6.999  & & 0.141 & & 1.28 & 30.8 \\
   & O I  &  7.063  & & 0.137 & & 1.83 & 43.9 \\
   & F I  &  7.119  & & 0.086 & & 2.26 & 58.1 \\
S  & He I &  5.039 & 5.039 & 0.711 & 0.767 & 6.23 & 6.23 \\
   &      &  4.300 \tablenotemark{b} & 4.299 & 0.158 & 0.153 & 1.90 & 2.01 \\
   & Li I &  5.084  & & 0.432 & & 5.57 & 22.7 \\
   &      &  5.086  & & 0.196 & & 5.05 & 7.68 \\
   & Be I &  5.126  & & 0.672 & & 5.68 & 14.5 \\
   &      &  4.441 \tablenotemark{b} & & 0.139 & & 1.56 & 15.8 \\
   & B I  &  5.176  & & 0.234 & & 5.82 & 11.4 \\
   &      &  5.183  & & 0.230 & & 2.86 & 25.1 \\
   & C I  &  5.234  & & 0.208 & & 1.69 & 35.7 \\
   &      &  5.222  & & 0.175 & & 1.42 & 17.0 \\
   & N I  &  5.276  & & 0.148 & & 2.36 & 41.2 \\
   & O I  &  5.320  & & 0.145 & & 3.42 & 52.8 \\
   & F I  &  5.359  & & 0.093 & & 4.30 & 65.5 \\
Ar & He I &  3.949 & 3.949 & 0.717 & 0.775 & 10.2 & 10.2 \\
   &      &  3.366 \tablenotemark{b} & 3.365 & 0.156 & 0.155 & 3.06 & 3.25 \\
   & Li I & 3.981  & & 0.444 & & 9.34 & 9.69 \\
   &      & 3.983  & & 0.192 & & 8.06 & 9.66 \\
   & Be I & 4.010  & & 0.681 & & 9.42 & 18.5 \\
   &      & 3.464 \tablenotemark{b} & & 0.140 & & 2.59 & 2.62 \\
   & B I  & 4.050  & & 0.250 & & 5.08 & 27.7 \\
   &      & 4.045  & & 0.244 & & 9.95 & 16.9 \\
   & C I  & 4.086  & & 0.236 & & 3.15 & 39.2 \\
   &      & 4.078  & & 0.175 & & 2.34 & 24.6 \\
   & N I  & 4.117  & & 0.153 & & 4.01 & 45.4 \\
   &      & 4.115  & & 0.103 & & 4.06 & 45.4 \\
   & O I  & 4.148  & & 0.151 & & 5.84 & 61.1 \\
   & F I  & 4.176  & & 0.098 & & 7.46 & 77.4 \\
Ca & He I & 3.177 & 3.177 & 0.717 & 0.782 & 15.8 & 15.8 \\
   &      & 2.706 \tablenotemark{b} & 2.705 & 0.154 & 0.155 & 4.68 & 4.97 \\
   & Li I & 3.200  & & 0.454 & & 14.8 & 15.0 \\
   &      & 3.202  & & 0.184 & & 12.0 & 14.0 \\
   & Be I & 3.221  & & 0.685 & & 14.7 & 24.1 \\
   &      & 2.776 \tablenotemark{b} & & 0.140 & & 4.03 & 4.08 \\
   & B I  & 3.250  & & 0.267 & & 8.43 & 31.1 \\
   &      & 3.247  & & 0.252 & & 16.0 & 24.7 \\
   & C I  & 3.276  & & 0.273 & & 5.65 & 43.0 \\
   &      & 3.270  & & 0.158 & & 3.29 & 34.6 \\
   & N I  & 3.301  & & 0.156 & & 6.38 & 50.1 \\
   &      & 3.299  & & 0.106 & & 6.51 & 50.1 \\
   & O I  & 3.323  & & 0.154 & & 9.28 & 70.0 \\
   & F I  & 3.344  & & 0.102 & & 12.1 & 90.5 \\
Fe & He I & 1.850 & 1.851 & 0.689 & 0.798, 0.704 \tablenotemark{c} & 44.7 & 44.7 \\
   &      & 1.573 \tablenotemark{b} & 1.575, 1.573 \tablenotemark{c}& 0.144 &
   0.156, 0.138 \tablenotemark{c}& 13.0 & 13.8 \\
   & Li I & 1.861 & 1.860 \tablenotemark{d} & 0.469 & 0.491 \tablenotemark{d} & 45.1 & 45.2 \\
   &      & 1.864 & 1.864 \tablenotemark{d} & 0.147 & 0.146 \tablenotemark{d} & 28.3 & 32.4 \\
   & Be I & 1.870  & & 0.666 & & 42.3 & 52.6 \\
   &      & 1.605 \tablenotemark{b} & & 0.134 & & 11.6 & 11.8 \\
   & B I  & 1.883  & & 0.300 & & 28.2 & 52.6 \\
   &      & 1.883  & & 0.259 & & 48.7 & 65.2 \\
   & C I  & 1.895  & & 0.397 & & 24.6 & 64.4 \\
   & N I  & 1.908  & & 0.154 & & 18.8 & 69.7 \\
   &      & 1.906  & & 0.119 & & 21.8 & 72.1 \\
   & O I  & 1.918  & & 0.151 & & 27.4 & 106.2 \\
   & F I  & 1.927  & & 0.109 & & 21.2 & 144.0 \\

\enddata
%\tablenotetext{a}{ Drake 1988}
\tablenotetext{a}{ Primarily \citet {verner96} }
\tablenotetext{b}{ 1s - 3p transitions }
\tablenotetext{c}{ NIST database
(http://physics.nist.gov/cgi-bin/AtData/main\_asd); quoted only where available
and different from \citet {verner96} }
\tablenotetext{d}{ \citet {nahar01}}
\end{deluxetable}

\clearpage

\begin{figure}
%  \plotone{f1.eps}
  \plotone{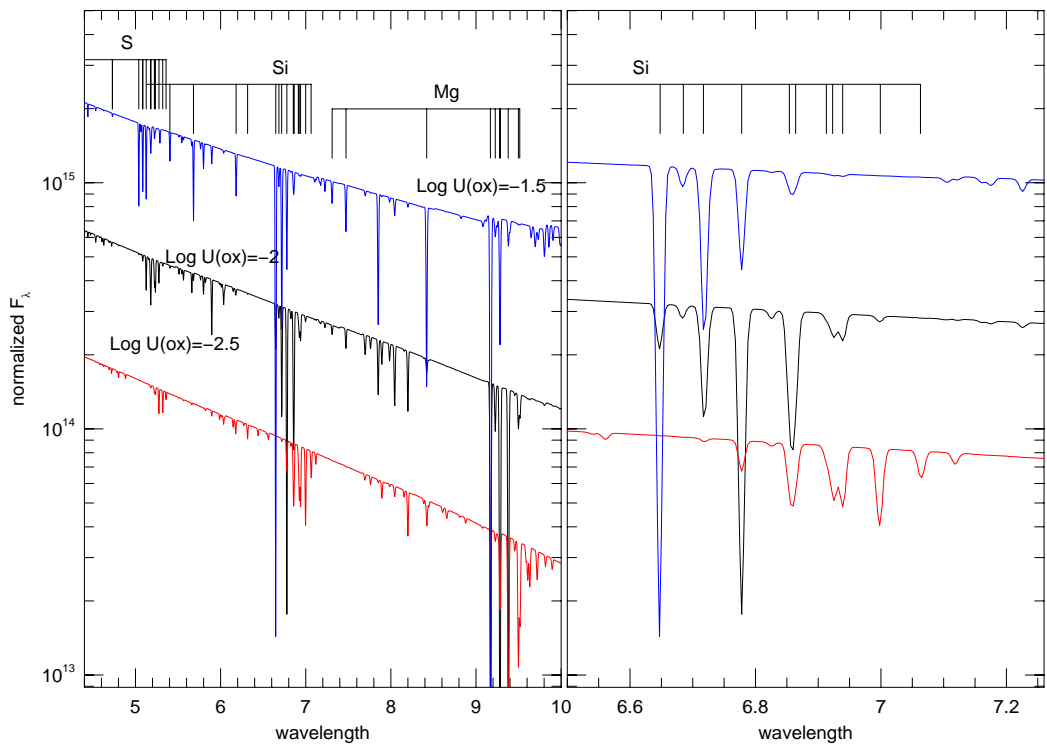}
  \rotate{}
%  \label{three-Ux}
%  \centerline{\psfig{figure=f1.eps,angle=0,height=6in}}
  \caption{Magnesium, silicon, and sulfur 1s - $n$p absorption 
  lines for a low density photoionized gas illuminated by a
  power-law X-ray continuum with $\Gamma=1.8$, for three different ionization 
  parameters (\Uox) as indicated. A column density of 10$^{22}$ \cmii\ is 
  assumed. The rulers above the spectra mark the strongest absorption lines 
  for each of the H-like through F-like ions. 
  Left: The full 4.7 -- 10 \AA\ range. Right: The silicon complex enlarged.}
\end{figure}                

\clearpage

\begin{figure}
%  \plotone{f2.eps}
  \plotone{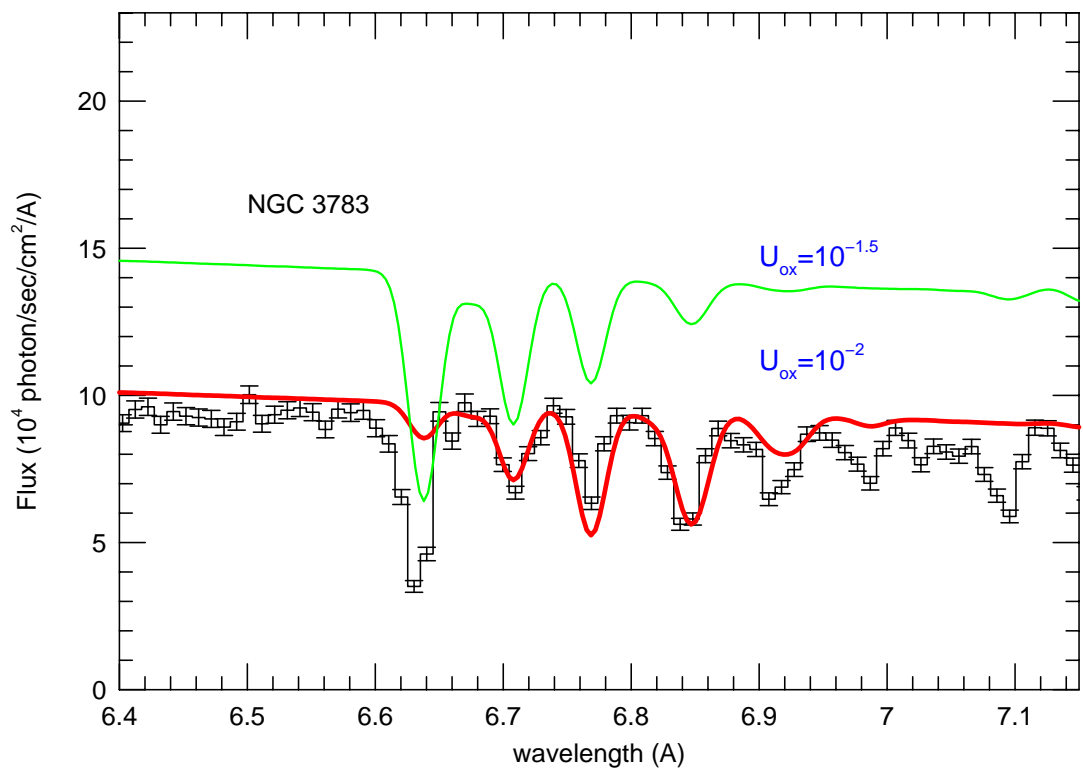}
  \rotate{}
%  \label{n3783_silicon}
%  \centerline{\psfig{figure=f2.eps,angle=0,height=6in}}
  \caption{Computed inner-shell silicon lines with a turbulent velocity of 
  300 km/s compared with the \hetgs\ observation of NGC 3783 \citep {kaspi01}.
  The \Uox=10$^{-2}$  model is normalized to the observed continuum and the 
  \Uox=10$^{-1.5}$ model is arbitrarily shifted for clarity of
  presentation. Each absorption trough in the spectrum corresponds to one
  charge state of Si (compare Table 1). The feature at $\sim$ 7.1 \AA\ is a Mg 
  line.}
\end{figure}                

\end{document}